# Scattering of a plasmonic nanoantenna embedded in a silicon waveguide.

Marta Castro-Lopez[*,1], Nuno de Sousa[2], Antonio Garcia-Martin[3], Frederic Y. Gardes[4], and Riccardo Sapienza[1].

[1] *Physics Department, King's College London, Strand, WC2R 2LS London, United Kingdom.*
[2] *Departamento de Fisica de la Materia Condensada, Universidad Autonoma de Madrid, 28049, Madrid, Spain,*
[3] *IMM-Instituto de Microelectronica de Madrid (CNM-CSIC), Isaac Newton 8, PTM, Tres Cantos, E-28760, Spain,*
[4] *Optoelectronics Research Centre (ORC), Building 53 Mountbatten, University of Southampton, SO17 1BJ Southampton, United Kingdom.*

[*]*Marta.Castro_Lopez @kcl.ac.uk*

**Abstract:** Plasmonic antennas integrated on silicon devices have large and yet unexplored potential for controlling and routing light signals. Here, we present theoretical calculations of a hybrid silicon-metallic system in which a single gold nanoantenna embedded in a single-mode silicon waveguide acts as a resonance-driven filter. As a consequence of scattering and interference, when the resonance condition of the antenna is met, the transmission drops by 85% in the resonant frequency band. Firstly, we study analytically the interaction between the propagating mode and the antenna by including radiative corrections to the scattering process and the polarization of the waveguide walls. Secondly, we find the configuration of maximum interaction and numerically simulate a realistic nanoantenna in a silicon waveguide. The numerical calculations show a large suppression of transmission and three times more scattering than absorption, consequent with the analytical model. The system we propose can be easily fabricated by standard silicon and plasmonic lithographic methods, making it promising as real component in future optoelectronic circuits.

**1. Introduction**

Silicon photonics is poised to become the technology of choice in integrated photonic circuits for optical interconnects, lab-on-a-chip processing and low power sensing networks [1-6]. Device optimization and maximal light-matter interaction are pushing the silicon technology to explore subwavelength architectures where optical processes can be strongly enhanced by nanophotonic design [7, 8]. Plasmonic integration is emerging as a promising platform [9, 10] reaching beyond light wavelength scales by means of metallic structures supporting localized surface plasmon polaritons (SPPs) [11]. The bound nature of SPPs at metal-dielectric interfaces leads to localization of propagating light waves into nanometer-scale volumes and a subsequent enhancement of most optical processes, e.g. harmonic conversion [12], light emission [13] and absorption [14]. Integrated on photonic chips, the strong field enhancement of plasmonic systems has been exploited for the realization of efficient hybrid metal-dielectric lasers [15, 16], high-speed and broadband modulators [17-19], enhanced photodetectors [20] or plasmonic light concentrators in silicon waveguides [21, 22].

The small footprint and unique ability of plasmonic antennas to mediate efficiently the interaction between near-field and far-field radiation [23, 24] make them inherently suitable as optical transducer for integrated photonic circuits. Hybrid geometries combining plasmonic antennas and cavities based on dielectric photonic crystals [25] and metallic mirrors [26] have shown a modulated optical response due to coupling between localize and stationary resonances.

The effect of a plasmonic antenna on a propagating optical mode inside a dielectric waveguide, however, is largely unexplored. Pioneering numerical calculations by Bruck et al. were focused on the coherent perfect absorption of two standing waves via external phase manipulation [27] and more recently, experiments from Peyskens et al. were dedicated to the control of antenna resonances by evanescence coupling to guided modes [28]. The simplified model of scattering of a dipole in a waveguide has instead been studied with the goal of maximizing the coupling of the light emitted by a single atom [29] or molecule [30] into the waveguide and in the context of radiation pressure experienced by a small particle [31].

Here we present a simple analytical model and numerical investigations of the scattering properties of a single plasmonic nanoantenna embedded in a single-mode silicon waveguide for operation at telecom wavelengths (Fig. 1). The nanoantenna is predicted to strongly interact with the propagating mode of the waveguide, and to reduce the waveguide transmission to 15% for wavelengths resonant with the antenna. This resonance-driven optical filtering can be tuned by changing the size of the antenna and is subwavelength as the antenna is only ~$\lambda/12$ large. Taking advantage of the resonant properties of plasmonic antennas and of the low-loss light guiding of silicon waveguides, the hybrid systems could be integrated into photonic chips for the realization of optical functionalities with ultrafast response and ultrasmall footprint.

**2. Methods**

We study here a metallic nanoantenna placed inside a silicon single-mode waveguide (Fig. 1 (a)). The antenna is a gold nanorod with 40 nm thickness, 40 nm height and length varying from 10 to 350 nm so the first order resonance mode lays on the spectral window of interest ($\lambda$ in the range 1-2.2 μm).

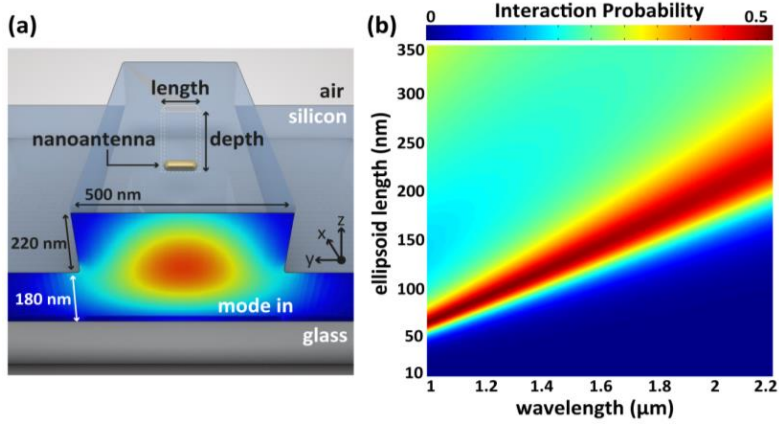

**Figure 1. Single plasmonic nanoantenna embedded in a silicon waveguide.** (a) 3D representation of the system under study. A gold nanorod is inserted in a gap created in a single-mode silicon waveguide. A TE mode is lunched from one side of the waveguide and interacts with the metallic nanoantenna. (b) Total interaction between the ellipsoid nanoantenna and the waveguide mode at different wavelengths for increasing size of the long axis of the ellipsoid (from 10 to 350 nm). The maximum interaction happens at the resonance length of the ellipsoid and reaches values of ~55%.

We study analytically and numerically the interaction between this type of antenna and the propagating mode while analyzing in detail the configuration of maximum interaction. Multiple fabrication methods can be envisioned to position the antenna within the waveguide; a self-aligned method where the hole and the ridge are etched at the same time, an oxide hard mask where the waveguide and the antenna recess are first defined and subsequently etched in turns or simply two lithography steps using e-beam where again the waveguide and the antenna recess are etched in turns. Subsequently to the silicon etching, the gold nanoantenna can be placed inside it using a lift off process.

**Analytical Model**

Scattering theory can describe light scattering from a plasmonic nanoantenna in a homogeneous medium, instead scattering in a waveguide requires more care. In the homogenous medium the antenna can be modeled as a classical point dipole with the polarizability corresponding to a prolate ellipsoid $\alpha$ [32]. The extinction cross-section $\sigma_{ext}$ of such a dipole in silicon is:

$$\sigma_{ext} = \frac{k}{\varepsilon_0 \varepsilon_m |\mathbf{E}_0|^2} \Im \left\{ \mathbf{E}_0^* \cdot \mathbf{p} \right\}$$

where $\mathbf{p}$ is the dipole, $\mathbf{E}_0$ is the incoming field, and the dielectric constant of silicon is $\varepsilon_m = 12.01$. On-resonance values of $\sigma_{ext}$ are calculated to be in the range 0.03-0.035 $\lambda^2$. The presence of the waveguide walls can be taken into account by considering the appearance of image dipoles, which interfere with the driving dipole. This self-interaction can be included in the model by renormalizing the antenna dipole in the homogenous medium [33] and calculating the new $\mathbf{p}$ via

$$\mathbf{p} = \left[ \mathbb{I} - k^2 \alpha \left\{ \mathbf{G}(\mathbf{r}_p, \mathbf{r}_{im1}) + \mathbf{G}(\mathbf{r}_p, \mathbf{r}_{im2}) \right\} \left\{ \frac{\varepsilon_{Si} - 1}{\varepsilon_{Si} + 1} \boldsymbol{\eta} \right\} \right]^{-1} \mathbf{p}_0$$

where $\mathbf{p_0}$ is the non-interacting dipole, $\alpha$ the polarizability in silicon, $\mathbf{G}$ is the Green tensor connecting the images and the antenna, $k$ the wave vector and $\boldsymbol{\eta}$ is a matrix that takes into account the reflection plane (see Appendix). Here we calculate the contribution to first order (i.e., including in the dressed dipole only the first image dipole on both sides of the waveguide) and within a perturbative approach based on a quasistatic limit, which means that the optical theorem is not strictly fulfilled.

We model the waveguide mode as a Gaussian function of waist ~ λ/3 [34] and define the interaction probability between the mode and the antenna as the integral of the waveguide mode over the area represented by the extinction cross-section of the antenna. This gives us the interaction probability values plotted in Fig. 1 (b). The antenna resonance is visible as a maximum of interaction that shifts to larger wavelengths when the ellipsoid major axis increases. Both the antenna dipole moment (and therefore also $\sigma_{ext}$) and the mode size of the single-mode waveguide increase for larger wavelengths, leading to on-resonance values of the interaction larger than 0.5 for all wavelengths (~ 50% interaction).

From the calculations, we find that ~ 24% of the light is absorbed in the metal while ~ 76% is scattered (i.e., $\sigma_{abs} = 1.99\times10^4$ nm$^2$ and $\sigma_{sca} = 6.28\times10^4$ nm$^2$ for a resonant ellipsoid at telecom wavelengths). Note that the obtained ratio scattering/absorption ≈ 3 is only attainable if the radiative corrections to the polarizability [35, 36] are taken into account (see Appendix). This correction is largely ignored in the literature whenever absorption is present in the particle, but when the same particle is embedded in a high index medium the scattering can be dominant even in presence of absorption.

From the design point of view, maximal extinction is not achieved for maximal interaction, as the low dimensionality of the system implies that interference between the transmitted and scattered light, which is phase-shifted upon scattering, will be very effective. Moreover, 100% interaction will lead to some non-zero transmission due to the scattered light propagating through the waveguide.

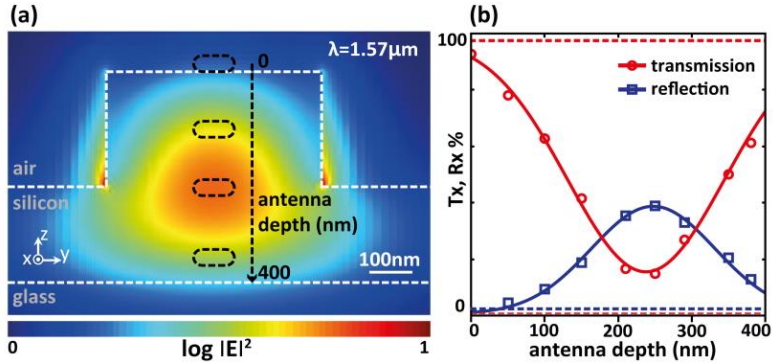

**Figure 2. Maximum optical interaction between nanoantenna and propagating mode.** (a) FDTD intensity distribution map of the propagating mode without the antenna at a wavelength of λ= 1.57 μm. (b) Percentage of the transmitted (red circles) and reflected (blue squares) power as a function of the position (depth) of the nanoantenna (shown with black dashed lines in (a)). The dashed lines represent transmission (red) and reflection (blue) of the mode without the presence of the antenna. The position of maximum interaction is found to be at a depth of 250 nm inside the waveguide.

**Numerical model**

Interference effects on the dipole moment as well as in the transmitted and scattered light, in a realistic 3D system, can be studied with numerical calculations which include the full electromagnetic interaction as well as the fine deviation in shape of the antenna from an ellipsoid and the small air gap we expect over the antenna if fabricated by electron-beam lithography. The simulations are performed using a FDTD method (Lumerical Solutions, Inc.) and are based on the structure depicted in Fig. 1 (a). A T-inverted-shape silicon waveguide with a ridge of 500 nm in width and 220 nm in height is placed on 180 nm of silicon with a glass substrate underneath. The modes of the waveguide are calculated using a mode solver.

This specific waveguide dimension supports single-mode propagation for TE polarization at telecom wavelengths of λ = 1.55 μm and in particular in the range 1-2.2 μm investigated here. For similar waveguide design, the measured transmission losses have been demon-strated to be better than 1 dB/cm [37] when fabricated with advanced fabrication tools such as immersion lithography. The nanoantenna presents a first order resonant mode at a wavelength of λ = 1.57 μm and its dimensions are 106×40×40 nm (length, width and height respectively). The hole is filled with air and it matches the sizes of the length and the width of the nanorod.

The system is excited by injecting the calculated first order TE mode into the waveguide. The nanoantenna is placed with its long y-oriented axis perpendicular to the direction of propagation (x-axis). The depth of the hole and therefore the position of the nanoantenna, is varied from 0 nm (nanorod on the surface of waveguide) to 400 nm (nanorod on top of the glass substrate surrounded by silicon) crossing the waveguide from the surface of the ridge to the substrate. The interaction between the propagating mode and the nanoantenna is studied by placing multiple transmission plane monitors before and after its position. The scattering from the antenna is recorded with monitors around the hole (on top, on the sides and below the waveguide). The refractive indexes of silicon and glass are taken from Palik [38] while the refractive index of gold is taken from Johnson and Christy [39].

## 3. Results and Discussion

The depth of the hole has been varied in order to achieve the maximum interaction between the propagating mode and the antenna. Fig. 2 (a) shows the intensity profile of the TE mode inside the waveguide (marked with white dashed lines). The mode is confined inside the ridge and shows a maximum around the center of the waveguide. Fig. 2 (b) shows the transmission and reflection values for different antenna depths (marked with black dashed lines in Fig. 2 (a)). The red circles and blue squares represent calculation points for the transmission and reflection respectively while the solid lines are guides for the eye. The dashed lines show the transmission (red) and reflection (blue) of the system without the antenna.

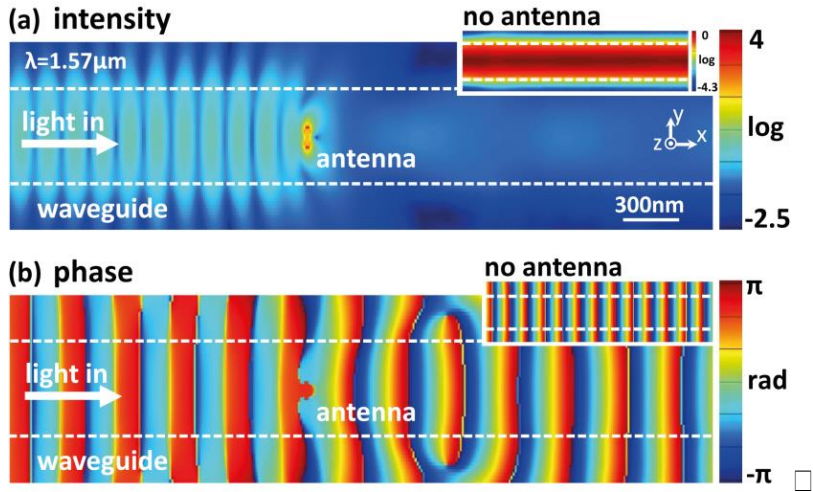

**Figure 3. Optical response of the hybrid metal-silicon system.** FDTD distribution maps of the (a) intensity and (b) phase of a TE mode propagating (from left to right) through the silicon waveguide and interacting with the metallic antenna at its resonance wavelength (λ=1.57μm). Light is almost completely blocked by the antenna and partially reflected backwards creating an interference pattern. The antenna also adds a phase shift to the propagating mode. The insets show the intensity (up) and phase (down) of the propagating TE mode without the presence of the antenna.

While the transmission through the waveguide is almost unaffected when the nanoantenna is on the surface of the waveguide (depth = 0 nm), when the antenna is lowered, it drops continuously until the antenna reaches the middle of the waveguide (depth = 250 nm) to increase again when it reaches the bottom of the waveguide (depth = 400 nm). As expected, the position of maximum interaction (depth = 250 nm) corresponds also to the position of the maximum of the mode. At that depth, the symmetry and polarization of the waveguide mode matches that of the dipolar mode of the antenna and maximal overlap is achieved.

Next we focus on the case of maximal interaction, i.e. for the nanoantenna in a hole at a depth of 250 nm, and we record the intensity and phase of the propagating light in the plane of the antenna. Fig. 3 (a) shows the intensity distribution map (in logarithmic scale) when the mode propagates from left to right and how its propagation is affected by the presence of the antenna. The antenna blocks a large fraction (~85%) of the transmission while reflecting part of the intensity backwards which creates an interference pattern with the incoming field, visible in Fig 3 (a). As a comparison, the inset shows a clear uniform propagation for the case without the antenna. The phase distribution map (Fig. 3 (b)) also confirms that the antenna imprints a phase shift to the transmitted signal. The inset for the case of waveguide without antenna shows the usual plane-wave like phase evolution of the waveguide mode.

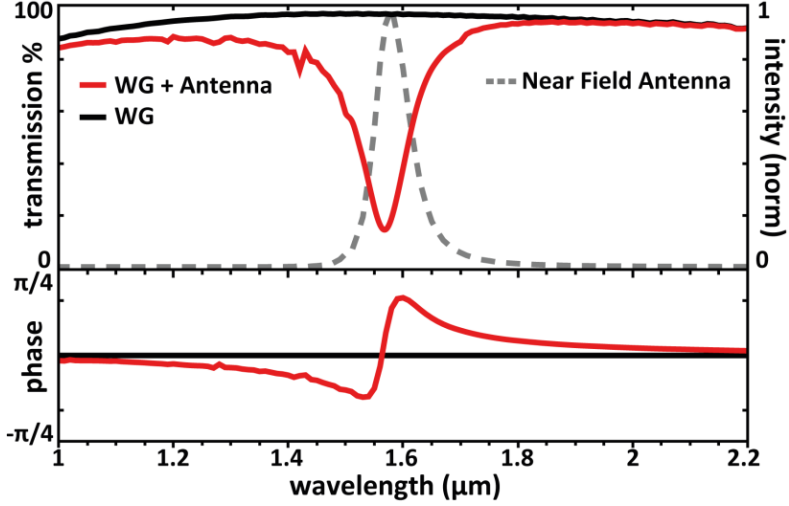

**Figure 4. Spectral response of the hybrid metal-silicon system.** Power transmission through the silicon waveguide (top) and spectral phase (bottom) of the system with (red line) and without (black line) the metallic antenna. At the resonance wavelength of the antenna, the transmission is blocked around 85% and the phase shifts almost $\pi/3$ rad when the antenna is present. The gray dashed curves show the normalized near-field intensity (top) and phase (bottom) in the vicinity of the antenna that corresponds to that of a harmonic oscillator.

In order to quantify the effect of the interaction with the antenna, in Fig. 4 we plot the transmission and phase as a function of the wavelength (top and bottom panels respectively) recorded 4 μm after the position of the antenna. The red and black solid lines show respectively the cases with and without the antenna. When the antenna is present (red line), a large dip in the transmission (Fig. 4 top panel) centered at telecom wavelengths ($\lambda = 1.55$ μm) is visible. Both shape and position of the dip match the spectral resonance of the antenna (gray dashed line, recorded as the near field intensity inside the air gap, 3 nm away from the antenna surface). At the position of the resonance ($\lambda = 1.57$ μm), the propagating mode is almost completely blocked by the nanoantenna, which reduces the power transmission from 98% (waveguide alone) to 15% (waveguide plus antenna).

The phase response of the system is shown in Fig. 4 bottom panel. While the phase of a TE mode propagating through the waveguide (black line) is spectrally flat (all wavelengths propagate in phase when the material intrinsic dispersion is neglected), the presence of the antenna adds a phase shift between the lower and higher wavelengths around the resonance wavelength. Our simulations show a phase jump, recorded 4 μm after the antenna, (red line) equal to $\Delta\varphi_{wg+ant} \approx 0.3\pi$. The smooth shape of the phase jump is consistent with the fact that our metallic antenna behaves like a damped harmonic oscillator due to inherent ohmic absorption and scattering losses.

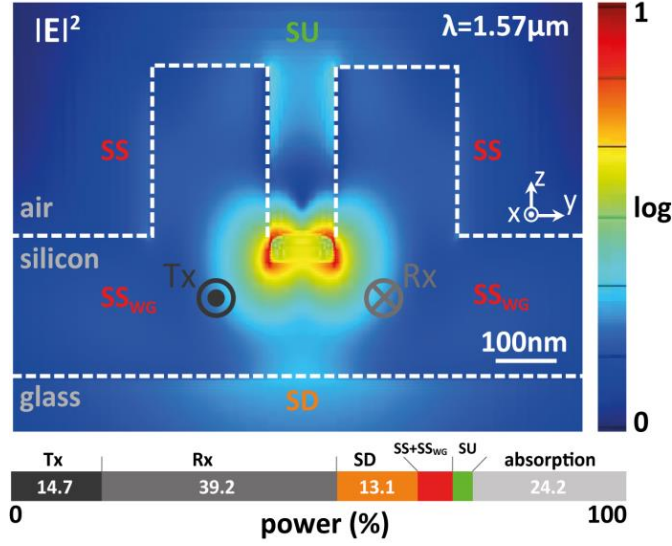

**Figure 5. Intensity distribution and power dissipation channels**. (Top) Intensity distribution map at the plane of the antenna showing the localization of the fields inside the waveguide. (Bottom) A detailed monitoring of the power dissipation channels reveals around 15% transmission (Tx), 39% reflection (Rx), 13% scattering towards the substrate (SD), 6% scattering towards the sides (SS+$SS_{WG}$) and 3% scattering up (SU) which leaves around 24% accounting for the losses of the waveguide and the absorption in the metal.

Finally, out of the 85% of the light blocked by the antenna, 40% is reflected backwards in the waveguide mode. In order to get a full understanding of the interaction process, we monitored the power scattered in all directions surrounding the antenna. Fig. 5 (top panel) shows the intensity distribution map in a XZ plane at the position of the antenna (y = 0 nm) for the wavelength of resonance. The bottom panel shows a color bar representing the 100% of the power injected into the waveguide and its dissipation channels. 14.69% of the power is transmitted (+y direction) and 39.20% is reflected (-y direction). From the remaining 46%, the 13.14% is emitted towards the glass substrate (scattering down, SD), 2.81% is emitted out through each side of the waveguide (scattering sides, SS+$SS_{WG}$) and 3.11% is scattered up towards the air (scattering up, SU). The rest 24.24% is lost and therefore accounts for the losses of the waveguide and the absorption of the metal. Since we also monitored the transmission through the waveguide without the antenna, we can calculate the percentage of light lost in the waveguide as $WG_{losses}$ = 1 - $P_{mode\_out}$/ $P_{mode\_in}$ where $P_{mode\_in}$ is the power of the mode injected into the waveguide and $P_{mode\_out}$ is the power of the mode arriving at the end of the waveguide. We calculate the waveguide losses to be 2.29% that sets the absorption of the metal at 21.95%.

We can compare the numerical results with the analytical model of an ellipsoidal antenna in a waveguide that we introduced earlier. An ellipsoid with length resonant at λ = 1.57 μm was predicted to interact with the 50% of the propagating wave by scattering 76% and absorbing 24% (i.e., 38% and 12% of the total light, respectively). The larger absorption obtained with the FDTD simulation can be attributed to the volume difference between the nanorod and the ellipsoid (a factor of 1.5). These values are compatible with the analytical calculations presented in first section, when radiative corrections to the polarizability are included.

## 4. Conclusion

In conclusion, this hybrid silicon-metallic system acts as an ultrasmall resonance-driven optical filter with transmission extinction down to 15% in a ~90 nm bandwidth. The position of the dip in transmission depends on the dimensions of the antenna and the refractive index surrounding it. Therefore, it could be tuned to any spectral position by changing the antenna length or by filling the gap with different materials. Moreover, the active tuning of the resonance position of the antenna could be achieved by pumping with an ultrafast laser pulse as shown in [40] and could lead to the realization of an ultrafast optical switch or tunable filter based on this design. The bandwidth of operation of this filter could also be narrowed or broadened by using different antenna geometries (for example supporting Fano resonances [41]), or using different metals (as for example silver or aluminum). As a future development, in a multimode waveguide, a specially designed nanoantenna could enable mode conversion or mode filtering by scattering interference. The subwavelength footprint, high performance and simplicity of plasmonic antennas could be a key asset for future optical circuits with impact in very diverse areas, ranging from telecom applications to quantum computation and sensing.

## Appendix

*Scattering cross sections*

In this section we explicitly write the extinction, scattering and absorption cross sections as a function of the response dipole of a particle that is immersed in a dielectric medium $\varepsilon_m$, instead of the usual expressions that are referred to the polarizability of the particle. This is convenient when that dipole response is convenient when that dipole response is affected by the environment. The extinction cross section is given by:

$$\sigma_{ext} = \frac{k}{\varepsilon_0 \varepsilon_m |\mathbf{E}_0|^2} \Im \{\mathbf{E}_0^* \cdot \mathbf{p}\}$$

where $k$ is the wave vector of the incident light in the medium and $\mathbf{E_0}$ is the incoming external wave (usually a plane wave) at the position of the particle [35]. The absorption cross section is defined as:

$$\sigma_{abs} = \frac{k}{\varepsilon_0 \varepsilon_m |\mathbf{E}_0|^2} \left\{ \Im (\mathbf{p} \cdot \mathbf{E}^*) - \frac{k^3}{6\pi \varepsilon_0 \varepsilon_m} |\mathbf{p}|^2 \right\}$$

where $\mathbf{E}$ is the total field at the position of the particle, i.e. it contains scattering effects from the environment. Finally, the scattering cross section is given by:

$$\sigma_{sca} = \frac{k^4}{\varepsilon_0^2 \varepsilon_m^2 6\pi |\mathbf{E}_0|^2} |\mathbf{p}|^2$$

The link with the usual expressions in terms of the polarizability are readily obtained using:

$$\mathbf{p} = \varepsilon_0 \varepsilon_m \alpha \mathbf{E}$$

where $\mathbf{E} = \mathbf{E_0}$ for an isolated particle.

*Polarizability of an elongated particle in a medium*

In order to have a self-contained description, let us consider here expression of the polarizability of a particle in a medium. The static polarizability of a particle whose dielectric is $\varepsilon_p$ in a $\varepsilon_m$ reads as:

$$\alpha_{0,ii} = 3\nu \frac{\varepsilon_p - \varepsilon_m}{3\varepsilon_m + 3L_i(\varepsilon_p - \varepsilon_m)}$$

where $\nu$ is the volume of the particle and the geometrical factors $L_i$ can be found in [32]. To ensure the optical theorem is fulfilled, it is necessary to consider the radiative correction to the static polarizability [35, 36]:

$$\alpha = \frac{\alpha_0}{1 - i\frac{k^3}{6\pi}\alpha_0}$$

*Scattering with waveguide walls*

In our system the particle lies inside a waveguide with $\varepsilon_m = \varepsilon_{Si} = 12.01$, in order to have a glimpse of the effect of the scattered fields going to the walls and back, we can use, as a zeroth-order approximation, the image dipoles induced on the outside of the waveguide. A further approximation is to consider the quasi-static case, but it is important to remember that this implicitly implies that the optical theorem is not fulfilled. In our case the deviation will be small, but a generalization must be taken with care. Let us consider a waveguide where the propagation is on the x-direction and the confinement is on the y-direction (the guide is infinite in the z-direction), where a dipolar metallic particle is located at its center. Within the mentioned framework, the image dipole takes the form [33]:

$$\mathbf{p}_{im} = \frac{\epsilon_w(\omega) - 1}{\epsilon_w(\omega) + 1}\eta\mathbf{p}$$

where $\boldsymbol{\eta}$ is the transformation matrix that represents the reflection over an axis. In our case (propagation along x-axis, and images along y-direction, i.e. zx plane):

$$\eta = \begin{bmatrix} -1 & 0 & 0 \\ 0 & 1 & 0 \\ 0 & 0 & -1 \end{bmatrix}$$

The electric field experienced by the particle is then given by (only two image dipoles):

$$\mathbf{E}(\mathbf{r}_p) = \mathbf{E}_0(\mathbf{r}_p) + \frac{k^2}{\epsilon_0 \epsilon_m}\{\mathbf{G}(\mathbf{r}_p, \mathbf{r}_{im1})\mathbf{p}_{im1} + \mathbf{G}(\mathbf{r}_p, \mathbf{r}_{im2})\mathbf{p}_{im2}\}$$

Which is reflected in a "dressed" polarization **p**:

$$\mathbf{p} = \left[\mathbb{I} - k^2\alpha\{\mathbf{G}(\mathbf{r}_p, \mathbf{r}_{im1}) + \mathbf{G}(\mathbf{r}_p, \mathbf{r}_{im2})\}\left\{\frac{\varepsilon_{Si} - 1}{\varepsilon_{Si} + 1}\eta\right\}\right]^{-1}\mathbf{p}_0$$

where $\mathbf{p}_0$ is the dipole without the action of the confinement walls.

**Acknowledgments**

We thank P. de Roque for discussions on the FDTD simulations. M.C-L. and R.S. acknowledge funding from the Engineering and Physical Sciences Research Council (EPSRC), FP7 EU Project People and a Leverhulme Trust Research Grant. N.S. acknowledges funding from the Spanish Ministry of Economy and Competitiveness through grant "MINIELPHO" FIS2012-36113-C03. The data is publicly available in Figshare.